\documentclass{article}
\usepackage{amssymb}
\usepackage{amsfonts}
\usepackage{amsmath}

\input{tcilatex}

\begin{document}

\begin{center}
\bigskip

{\large Perturbative Confinement in a 4-d Lorentzian Complex Structure
Dependent YM-like Model}

{\large \ }\bigskip

C. N. RAGIADAKOS

Pedagogical Institute

Mesogion 396, Agia Paraskevi, TK 15341, Greece

E-mail: crag@pi-schools.gr

\bigskip

\textbf{ABSTRACT}

I continue the study of a renormalizable four-dimensional generally
covariant Yang-Mills-like action, which depends on the Lorentzian complex
structure of spacetime and not its metric. The field equations and their
integrability conditions are written down explicitly. The model is studied
with the presence of two static external sources in the trivial cylindrical
complex structure. The energy of two static "colored" sources is found to
increase linearly with respect to their distance, providing an explicit
proof of their perturbative confinement. In the present model, confinement
is not a concequence of the non-Abelian character of the gauge group, but it
is implied by the complex structure dependence of the model.
\end{center}

\renewcommand{\theequation}{\arabic{section}.\arabic{equation}}

\section{INTRODUCTION}

\setcounter{equation}{0}

There is a strong experimental evidence that hadrons contain fundamental
constituents (the quarks) which cannot exist free. The spectrum of a linear
attractive potential between the quarks fits quite well the observed heavy
hadrons. In the context of the Standard Model, current strong interactions
are incorporated through an $SU(3)$\ Yang-Mills (YM) field which couples
with colored quarks. It is well known that the static potential of a YM
field is Coulomb like ($\frac{1}{r}$). It is generally believed that the
non-abelian YM Lagrangian somehow generates a linear potential, while no
explicit theoretical proof has yet been presented. This belief is supported
by the asymptotic freedom in the ultraviolet limit and some computer
calculations on the lattice. I will not continue on the current hadronic
phenomenology, the achievements of Quantum Chromodynamics and its failures.
The purpose of the present work is to provide an explicit counter example to
the general belief that in four dimensions a linear perturbative potential
is generated only in the context of Lagrangians with higher order
derivatives.

The present model emerged from my attempt to transfer in four dimensions the
characteristic property of the two-dimensional string action%
\begin{equation}
I_{S}=\frac{1}{2}\int d^{2}\!\xi \ \sqrt{-\gamma }\ \gamma ^{\alpha \beta }\
\partial _{\alpha }X^{\mu }\partial _{\beta }X^{\nu }\eta _{\mu \nu }
\label{i1}
\end{equation}%
to depend on the complex structure of the two-dimensional surface and not
its metric $\gamma _{\alpha \beta }$. It is well known that in the structure
coordinates $(z^{0},\ z^{\widetilde{0}})$ the string action takes the form%
\begin{equation}
I_{S}=\int d^{2}\!z\ (\partial _{0}X^{\mu })(\partial _{\widetilde{0}}X^{\nu
})\eta _{\mu \nu }  \label{i2}
\end{equation}%
which does not depend on the metric of the 2-dimensional surface.

The null tetrad form of the present model action\cite{RAG1999},\cite%
{RAG2008a} is 
\begin{equation}
\begin{array}{l}
I_{G}=\int d^{4}\!x\ \sqrt{-g}\ \left\{ \left( \ell ^{\mu }m^{\rho
}F_{\!j\mu \rho }\right) \left( n^{\nu }\overline{m}^{\sigma }F_{\!j\nu
\sigma }\right) +\left( \ell ^{\mu }\overline{m}^{\rho }F_{\!j\mu \rho
}\right) \left( n^{\nu }m^{\sigma }F_{\!j\nu \sigma }\right) \right\}  \\ 
\\ 
F_{j\mu \nu }=\partial _{\mu }A_{j\nu }-\partial _{\nu }A_{j\mu }-\gamma
\,f_{jik}A_{i\mu }A_{k\nu }%
\end{array}
\label{i3}
\end{equation}%
where $A_{j\mu }$ is a gauge field and $(\ell _{\mu },\,n_{\mu },\,m_{\mu
},\,\overline{m}_{\mu })$ is a null tetrad, which defines an integrable
complex structure\cite{FLAHE1974},\cite{FLAHE1976}. The metric tensor $%
g_{\mu \nu }$ and the complex structure tensor $J_{\mu }^{\;\nu }$ take the
form

\begin{equation}
\begin{array}{l}
g_{\mu \nu }=\ell _{\mu }n_{\nu }+n_{\mu }\ell _{\nu }-m_{{}\mu }\overline{m}%
_{\nu }-\overline{m}_{\mu }m_{\nu } \\ 
\\ 
J_{\mu }^{\;\nu }=i(\ell _{\mu }n^{\nu }-n_{\mu }\ell ^{\nu }-m_{\mu }%
\overline{m}^{\nu }+\overline{m}_{\mu }m^{\nu })%
\end{array}
\label{i4}
\end{equation}%
The integrability condition of this complex structure implies the Frobenius
integrability conditions of the pairs $(\ell _{\mu },\,\,m_{\mu })$ and $%
(n_{\mu },\,\overline{m}_{\mu })$. That is

\begin{equation}
\begin{array}{l}
(\ell ^{\mu }m^{\nu }-\ell ^{\nu }m^{\mu })(\partial _{\mu }\ell _{\nu
})=0\;\;\;\;,\;\;\;\;(\ell ^{\mu }m^{\nu }-\ell ^{\nu }m^{\mu })(\partial
_{\mu }m_{\nu })=0 \\ 
\\ 
(n^{\mu }m^{\nu }-n^{\nu }m^{\mu })(\partial _{\mu }n_{\nu
})=0\;\;\;\;,\;\;\;\;(n^{\mu }m^{\nu }-n^{\nu }m^{\mu })(\partial _{\mu
}m_{\nu })=0%
\end{array}
\label{i5}
\end{equation}%
which restricts the spacetime to have two geodetic and shear free
congruences.

Then Frobenius theorem states that there are four complex functions $%
(z^{\alpha },\;z^{\widetilde{\alpha }})$,\ $\alpha =0,\ 1$ , such that

\begin{equation}
dz^{\alpha }=f_{\alpha }\ \ell _{\mu }dx^{\mu }+h_{\alpha }\ m_{\mu }dx^{\mu
}\;\;\;\;,\;\;\;dz^{\widetilde{\alpha }}=f_{\widetilde{\alpha }}\ n_{\mu
}dx^{\mu }+h_{\widetilde{\alpha }}\ \overline{m}_{\mu }dx^{\mu }\;
\label{i6}
\end{equation}%
These four functions are the structure coordinates of the (integrable)
complex structure. Recall that in the euclidean manifolds the complex
structure is defined as a real tensor. But in the present case of Lorentzian
spacetimes the coordinates $z^{\widetilde{\alpha }}$ are not complex
conjugate of $z^{\alpha }$, because $J_{\mu }^{\;\nu }$ is no longer a real
tensor.

The difference between the present action and the ordinary Yang-Mills action
becomes more clear in the following covariant form of the action. 
\begin{equation}
I_{G}=-\frac{1}{8}\int d^{4}\!x\ \sqrt{-g}\ \left( 2g^{\mu \nu }\ g^{\rho
\sigma }-J^{\mu \nu }\ J^{\rho \sigma }-\overline{J^{\mu \nu }}\ \overline{%
J^{\rho \sigma }}\right) F_{\!j\mu \rho }F_{\!j\nu \sigma }  \label{i7}
\end{equation}%
where $g_{\mu \nu }$ is a metric derived from the null tetrad and $J_{\mu
}^{\;\nu }$ is the tensor (\ref{i4}) of the integrable complex structure.

The integrability of the Lorentzian complex structure is essential, because
only for these spacetimes the action can take a metric independent form,
which assures its renormalizability\cite{RAG2008b}. When we transcribe it in
its structure coordinates, it takes the following form 
\begin{equation}
\begin{array}{l}
I_{G}=\int d^{4}\!z\ F_{\!j01}F_{\!j\widetilde{0}\widetilde{1}}+com.\text{\ }%
conj. \\ 
\\ 
F_{j_{ab}}=\partial _{a}A_{jb}-\partial _{a}A_{jb}-\gamma
\,f_{jik}A_{ia}A_{kb}%
\end{array}
\label{i8}
\end{equation}%
which is metric independent, analogous to the form (\ref{i2}) of the string
action. Therefore we have to implement the integrability conditions (\ref{i5}%
) using Lagrange multipliers 
\begin{equation}
\begin{array}{l}
I_{C}=\int d^{4}\!x\ \sqrt{-g}\{\phi _{0}(\ell ^{\mu }m^{\nu }-\ell ^{\nu
}m^{\mu })(\partial _{\mu }\ell _{\nu })+\phi _{1}(\ell ^{\mu }m^{\nu }-\ell
^{\nu }m^{\mu })(\partial _{\mu }m_{\nu })+ \\ 
\\ 
\qquad +\phi _{\widetilde{0}}(n^{\mu }\overline{m}^{\nu }-n^{\nu }\overline{m%
}^{\mu })(\partial _{\mu }n_{\nu })+\phi _{\widetilde{1}}(n^{\mu }\overline{m%
}^{\nu }-n^{\nu }\overline{m}^{\mu })(\partial _{\mu }\overline{m}_{\nu
})+c.conj.\}%
\end{array}
\label{i9}
\end{equation}%
The complete action $I=I_{G}+I_{C}$ is self-consistent and the usual
quantization techniques may be used\cite{RAG1990},\cite{RAG1991},\cite%
{RAG1992}. The renormalizability is the great value of the present model,
because if supersymmetry is not found in the current experiments,
superstrings have to be abandoned. The characteristic properties of the
present model appear to be very appealing, to provide a pathway to a "theory
of everything".

The model uses the Newman-Penrose null tetrad formalism, but the essential
calculations of the present paper will be trivial. In section II the
complete field equations and their integrability conditions are explicitly
written down. Their forms are quite complicated for readers non familiar
with the Newman-Penrose formalism. The unfamiliar readers may skip it. But
these equations permit the familiar reader to understand the emergence of an
"energy-momentum" tensor as integrability conditions of the field equations.
In fact this tensor coincides with that found considering the YM field in an
external time independent null tetrad.

An indication of perturbative "gluonic" confinement has been presented in my
previous works\cite{RAG1999},\cite{RAG2008a} through the computation of the
static "gluonic" potential in the trivial spherical complex structure
determined by the following (spherical) null tetrad in spherical coordinates 
$(t,\ r,\ \theta ,\ \varphi )$ 
\begin{equation}
\begin{array}{l}
\ell _{\mu }=\left( 1\ ,\ -1\ ,\ 0\ ,\ 0\right)  \\ 
n_{\mu }=\frac{1}{2}\ \left( 1\ ,\ 1\ ,\ 0\ ,\ 0\right)  \\ 
m_{\mu }=\frac{-r}{\sqrt{2}}\ \left( 0\ ,\ 0\ ,\ 1\ ,\ i\sin \theta \right) 
\end{array}
\label{i10}
\end{equation}%
with its contravariant coordinates 
\begin{equation}
\begin{array}{l}
\ell ^{\mu }=\left( 1\ ,\ 1\ ,\ 0\ ,\ 0\right)  \\ 
n^{\mu }=\frac{1}{2}\ \left( 1\ ,\ -1\ ,\ 0\ ,\ 0\right)  \\ 
m^{\mu }=\frac{1}{r\sqrt{2}}\ \left( 0\ ,\ 0\ ,\ 1\ ,\ \frac{i}{\sin \theta }%
\right) 
\end{array}
\label{i11}
\end{equation}

If we expand the gauge field into the null tetrad%
\begin{equation}
A_{j\mu }=B_{j1}\ell _{\mu }+B_{j2}n_{\mu }+\overline{B_{j}}m_{\mu }+B_{j}%
\overline{m}_{\mu }  \label{i12}
\end{equation}%
we find the gauge field components $B_{j1},\ B_{j2},\ B_{j}$. In the present
null tetrad, the conjugate momenta of $B_{j1},\ B_{j2}$ vanish, i.e. $%
P_{j1}=0=P_{j2}$. Therefore we must assume $B_{j1}=0=B_{j2}$.\ Assuming the
convenient gauge condition 
\begin{equation}
\overline{m}^{\nu }\partial _{\nu }\left( r\ \sin \theta \ m^{\mu }A_{j\mu
}\right) +m^{\nu }\partial _{\nu }\left( r\ \sin \theta \ m^{\mu }A_{j\mu
}\right) =0  \label{i13}
\end{equation}%
the field equation takes the form 
\begin{equation}
\left( \frac{\partial ^{2}}{\partial t^{2}}-\frac{\partial ^{2}}{\partial
r^{2}}\right) \left( rm^{\mu }A_{j\mu }\right) =\left[ source\right]
\label{i14}
\end{equation}%
which apparently implies a linear "gluonic" potential for the field variable 
$\left( rm^{\mu }A_{j\mu }\right) $. It is not enough. I have also to show
that this variable diagonalizes the energy of the model.

In the precise complex structure (null tetrad) the energy of the YM field is 
\begin{equation}
\begin{array}{l}
E=\dint T_{g\ 0}^{0}\sqrt{-g}d^{3}x= \\ 
\quad =\dint drd\theta d\varphi \sin \theta \lbrack \partial _{t}(rm^{\mu
}A_{j\mu })\partial _{t}(r\overline{m}^{\mu }A_{j\mu })+\partial
_{r}(rm^{\mu }A_{j\mu })\partial _{r}(r\overline{m}^{\mu }A_{j\mu })]%
\end{array}
\label{i15}
\end{equation}%
Notice that the \underline{dynamical variable} of the gauge field is $\left(
r\ m^{\mu }A_{j\mu }\right) $, because this form diagonalizes the energy.
From its above field equation, we see that this dynamical variable
apparently gives a linear classical static potential.

In section III, I will review the trivial computation of the energy of two
electric charges. This trivial computation is reviewed, in order to reveal
its similarities and differences with the energy of two "gluonic" charges,
which is subsequently computed in the same section.

\section{FIELD\ EQUATIONS\ AND\ INTEGRABILITY\ CONDITIONS}

\setcounter{equation}{0}

Variation of the action with respect to the gauge field $A_{j\mu }$ \bigskip
gives the field equations 
\begin{equation}
\begin{array}{l}
D_{\mu }\{\sqrt{-g}[(\ell ^{\mu }m^{\nu }-\ell ^{\nu }m^{\mu })(n^{\rho }%
\overline{m}^{\sigma }F_{j\rho \sigma })+(n^{\mu }\overline{m}^{\nu }-n^{\nu
}\overline{m}^{\mu })(\ell ^{\rho }m^{\sigma }F_{j\rho \sigma })+ \\ 
\\ 
\qquad +(\ell ^{\mu }\overline{m}^{\nu }-\ell ^{\nu }\overline{m}^{\mu
})(n^{\rho }m^{\sigma }F_{j\rho \sigma })+(n^{\mu }m^{\nu }-n^{\nu }m^{\mu
})(\ell ^{\rho }\overline{m}^{\sigma }F_{j\rho \sigma })]\}=0%
\end{array}
\label{fe1}
\end{equation}%
where $D_{\mu }=\delta _{\ell j}\partial _{\mu }+\gamma f_{\ell jk}A_{k\mu }$
is the gauge symmetry covariant derivative and $\gamma $ the coupling
constant. Multiplying with the null tetrad, these equations take the form 
\begin{equation}
\begin{array}{l}
m^{\mu }D_{\mu }(\ell \overline{m}F_{j})+\overline{m}^{\mu }D_{\mu }(\ell
mF_{j})+(\ell \overline{m}F_{j})[(\nabla _{\mu }m^{\mu })+(nm\partial \ell
)]+ \\ 
\qquad +(\ell mF_{j})[(\nabla _{\mu }\overline{m}^{\mu })+(n\overline{m}%
\partial \ell )]=0 \\ 
\\ 
m^{\mu }D_{\mu }(n\overline{m}F_{j})+\overline{m}^{\mu }D_{\mu }(nmF_{j})+(n%
\overline{m}F_{j})[(\nabla _{\mu }m^{\mu })+(\ell m\partial n)]+ \\ 
\qquad +(nmF_{j})[(\nabla _{\mu }\overline{m}^{\mu })+(\ell \overline{m}%
\partial n)]=0 \\ 
\\ 
\ell ^{\mu }D_{\mu }(nmF_{j})+n^{\mu }D_{\mu }(\ell
mF_{j})+(nmF_{j})[(\nabla _{\mu }\ell ^{\mu })+(\ell \overline{m}\partial
m)]+ \\ 
\qquad +(\ell mF_{j})[(\nabla _{\mu }n^{\mu })+(n\overline{m}\partial m)]=0
\\ 
\end{array}
\label{fe2}
\end{equation}

Variation of the action with respect to the Lagrange multipliers $\phi
_{0},\ \phi _{1},\ \phi _{\widetilde{0}},\ \phi _{\widetilde{1}}$ imply the
complex structure integrability conditions on the tetrad (\ref{i5}).
Variation of the action with respect to the tetrad gives PDEs on the
Lagrange multipliers. In order to preserve the relations between the
covariant and contravariant forms of the tetrad we will use the identities 
\begin{equation}
\begin{array}{l}
\delta e_{a}^{\mu }=e_{a}^{\lambda }[-n^{\mu }\delta \ell _{\lambda }-\ell
^{\mu }\delta n_{\lambda }+\overline{m}^{\mu }\delta m_{\lambda }+m^{\mu
}\delta \overline{m}_{\lambda }] \\ 
\\ 
\delta \sqrt{-g}=\sqrt{-g}[n^{\lambda }\delta \ell _{\lambda }+\ell
^{\lambda }\delta n_{\lambda }-\overline{m}^{\lambda }\delta m_{\lambda
}-m^{\lambda }\delta \overline{m}_{\lambda }]%
\end{array}
\label{fe3}
\end{equation}%
where we denote $(e_{\mu }^{0}=\ell _{\mu }\ ,\,\,e_{\mu }^{1}=m_{\mu })$
and $(e_{\mu }^{\widetilde{0}}=n_{\mu }\ ,\,\,e_{\mu }^{\widetilde{1}}=\,%
\overline{m}_{\mu })$.\ Variation with respect to $\ell _{\lambda }$ gives
the PDEs 
\begin{equation}
\begin{array}{l}
2\ell ^{\lambda }(nmF_{j})(n\overline{m}F_{j})+m^{\lambda }(\ell nF_{j})(n%
\overline{m}F_{j})+\overline{m}^{\lambda }(\ell nF_{j})(nmF_{j})= \\ 
\qquad =-\nabla _{\mu }\left[ \phi _{0}(\ell ^{\mu }m^{\lambda }-\ell
^{\lambda }m^{\mu })\right] -\nabla _{\mu }\left[ \overline{\phi _{0}}(\ell
^{\mu }\overline{m}^{\lambda }-\ell ^{\lambda }\overline{m}^{\mu })\right] -
\\ 
\qquad \qquad -\ell ^{\lambda }\left[ \phi _{0}(nm\partial \ell )+\overline{%
\phi _{0}}(n\overline{m}\partial \ell )\right] -m^{\lambda }\left[ \phi
_{0}(\ell n\partial \ell )+\phi _{1}(\ell n\partial m)\right] - \\ 
\qquad \qquad -\overline{m}^{\lambda }\left[ \overline{\phi _{0}}(\ell
n\partial \ell )+\overline{\phi _{1}}(\ell n\partial \overline{m})\right] 
\\ 
\end{array}
\label{fe4}
\end{equation}%
which take the tetrad form 
\begin{equation}
\begin{array}{l}
m^{\mu }\partial _{\mu }\phi _{0}+\overline{m}^{\mu }\partial _{\mu }%
\overline{\phi _{0}}+\phi _{0}[(\nabla _{\mu }m^{\mu })+(\ell m\partial
n)-(nm\partial \ell )]+ \\ 
\qquad +\overline{\phi _{0}}[(\nabla _{\mu }\overline{m}^{\mu })+(\ell 
\overline{m}\partial n)-(n\overline{m}\partial \ell )]+2(nmF_{j})(n\overline{%
m}F_{j})=0 \\ 
\\ 
\ell ^{\mu }\partial _{\mu }\phi _{0}+\phi _{0}[(\nabla _{\mu }\ell ^{\mu
})+(\ell m\partial \overline{m})+(\ell n\partial \ell )]+\phi _{1}(\ell
n\partial m)+ \\ 
\qquad +(\ell nF_{j})(n\overline{m}F_{j})=0 \\ 
\end{array}
\label{fe5}
\end{equation}%
Variation with respect to $n_{\lambda }$ gives the PDEs 
\begin{equation}
\begin{array}{l}
2n^{\lambda }(\ell mF_{j})(\ell \overline{m}F_{j})-m^{\lambda }(\ell
nF_{j})(\ell \overline{m}F_{j})-\overline{m}^{\lambda }(\ell nF_{j})(\ell
mF_{j})= \\ 
\qquad =-\nabla _{\mu }\left[ \phi _{\widetilde{0}}(n^{\mu }\overline{m}%
^{\lambda }-n^{\lambda }\overline{m}^{\mu })\right] -\nabla _{\mu }\left[ 
\overline{\phi _{\widetilde{0}}}(n^{\mu }m^{\lambda }-n^{\lambda }m^{\mu })%
\right] - \\ 
\qquad \qquad -n^{\lambda }\left[ \phi _{\widetilde{0}}(\ell \overline{m}%
\partial n)+\overline{\phi _{\widetilde{0}}}(\ell m\partial n)\right] +%
\overline{m}^{\lambda }\left[ \phi _{\widetilde{0}}(\ell n\partial n)+\phi _{%
\widetilde{1}}(\ell n\partial \overline{m})\right] + \\ 
\qquad \qquad +\overline{m}^{\lambda }\left[ \overline{\phi _{\widetilde{0}}}%
(\ell n\partial \ell )+\overline{\phi _{\widetilde{1}}}(\ell n\partial m)%
\right]  \\ 
\end{array}
\label{fe6}
\end{equation}%
which take the tetrad form 
\begin{equation}
\begin{array}{l}
\overline{m}^{\mu }\partial _{\mu }\phi _{\widetilde{0}}+m^{\mu }\partial
_{\mu }\overline{\phi _{\widetilde{0}}}+\phi _{\widetilde{0}}[(\nabla _{\mu }%
\overline{m}^{\mu })+(n\overline{m}\partial \ell )-(\ell \overline{m}%
\partial n)]+ \\ 
\qquad +\overline{\phi _{\widetilde{0}}}[(\nabla _{\mu }m^{\mu
})+(nm\partial \ell )-(\ell m\partial n)]-2(\ell mF_{j})(\ell \overline{m}%
F_{j})=0 \\ 
\\ 
n^{\mu }\partial _{\mu }\phi _{\widetilde{0}}+\phi _{\widetilde{0}}[(\nabla
_{\mu }n^{\mu })+(n\overline{m}\partial m)-(\ell n\partial n)]-\phi _{%
\widetilde{1}}(\ell n\partial \overline{m})- \\ 
\qquad -(\ell nF_{j})(\ell mF_{j})=0 \\ 
\end{array}
\label{fe7}
\end{equation}%
Variation with respect to $m_{\lambda }$ gives the PDEs 
\begin{equation}
\begin{array}{l}
\ell ^{\lambda }(m\overline{m}F_{j})(n\overline{m}F_{j})+n^{\lambda }(m%
\overline{m}F_{j})(\ell \overline{m}F_{j})-2m^{\lambda }(\ell \overline{m}%
F_{j})(n\overline{m}F_{j})= \\ 
\qquad =-\nabla _{\mu }\left[ \phi _{1}(\ell ^{\mu }m^{\lambda }-\ell
^{\lambda }m^{\mu })\right] -\nabla _{\mu }\left[ \overline{\phi _{%
\widetilde{1}}}(n^{\mu }m^{\lambda }-n^{\lambda }m^{\mu })\right] - \\ 
\qquad \qquad -\ell ^{\lambda }\left[ \phi _{0}(m\overline{m}\partial \ell
)+\phi _{1}(m\overline{m}\partial m)\right] -n^{\lambda }\left[ \overline{%
\phi _{\widetilde{0}}}(m\overline{m}\partial n)+\overline{\phi _{\widetilde{1%
}}}(m\overline{m}\partial m)\right] + \\ 
\qquad \qquad +m^{\lambda }\left[ \phi _{1}(\ell \overline{m}\partial m)+%
\overline{\phi _{\widetilde{1}}}(n\overline{m}\partial m)\right]  \\ 
\end{array}
\label{fe8}
\end{equation}%
which take the tetrad form 
\begin{equation}
\begin{array}{l}
m^{\mu }\partial _{\mu }\overline{\phi _{\widetilde{1}}}+\overline{\phi _{%
\widetilde{1}}}[(\nabla _{\mu }m^{\mu })+(nm\partial \ell )-(m\overline{m}%
\partial m)]-\overline{\phi _{\widetilde{0}}}(m\overline{m}\partial n)- \\ 
\qquad -(\ell \overline{m}F_{j})(m\overline{m}F_{j})=0 \\ 
\\ 
m^{\mu }\partial _{\mu }\phi _{1}+\phi _{1}[(\nabla _{\mu }m^{\mu })+(\ell
m\partial n)-(m\overline{m}\partial m)]-\phi _{0}(m\overline{m}\partial \ell
)- \\ 
\qquad -(n\overline{m}F_{j})(m\overline{m}F_{j})=0 \\ 
\\ 
\ell ^{\mu }\partial _{\mu }\phi _{1}+n^{\mu }\partial _{\mu }\overline{\phi
_{\widetilde{1}}}+\phi _{1}[(\nabla _{\mu }\ell ^{\mu })+(\ell m\partial 
\overline{m})-(\ell \overline{m}\partial m)]+ \\ 
\qquad +\overline{\phi _{\widetilde{1}}}[(\nabla _{\mu }n^{\mu
})+(nm\partial \overline{m})-(n\overline{m}\partial m)]-2(\ell \overline{m}%
F_{j})(n\overline{m}F_{j})=0 \\ 
\end{array}
\label{fe9}
\end{equation}%
In order to simplify the relations, I made the bracket notations like $%
(nm\partial \ell )\equiv $ $(n^{\mu }m^{\nu }-$ $n^{\nu }m^{\mu })\partial
_{\mu }\ell _{\nu }$ for the spin coefficients and like $(nmF_{j})\equiv
n^{\mu }m^{\nu }F_{j\mu \nu }\ $for the gauge field components.

On the other hand the $e_{\ \mu }^{a}$ field equations imply the four
conserved currents 
\begin{equation}
\begin{array}{l}
\nabla _{\lambda }\{\ell ^{\lambda }[2(nmF_{j})(n\overline{m}F_{j})+\phi
_{0}(nm\partial \ell )+\overline{\phi _{0}}(n\overline{m}\partial \ell )]+
\\ 
\qquad +m^{\lambda }[(\ell nF_{j})(n\overline{m}F_{j})+\phi _{0}(\ell
n\partial \ell )+\phi _{1}(\ell n\partial m)]+ \\ 
\qquad +\overline{m}^{\lambda }[(\ell nF_{j})(nmF_{j})+\overline{\phi _{0}}%
(\ell n\partial \ell )+\overline{\phi _{1}}(\ell n\partial \overline{m})]\}=0
\\ 
\\ 
\nabla _{\lambda }\{n^{\lambda }\left[ 2(\ell mF_{j})(\ell \overline{m}%
F_{j})+\phi _{\widetilde{0}}(\ell \overline{m}\partial n)+\overline{\phi _{%
\widetilde{0}}}(\ell m\partial n)\right] - \\ 
\qquad -\overline{m}^{\lambda }[(\ell nF_{j})(\ell mF_{j})+\phi _{\widetilde{%
0}}(\ell n\partial n)+\phi _{\widetilde{1}}(\ell n\partial \overline{m})]-
\\ 
\qquad -m^{\lambda }\left[ (\ell nF_{j})(\ell \overline{m}F_{j})+\overline{%
\phi _{\widetilde{0}}}(\ell n\partial n)+\overline{\phi _{\widetilde{1}}}%
(\ell n\partial m)\right] \}=0 \\ 
\\ 
\nabla _{\lambda }\{\ell ^{\lambda }[(m\overline{m}F_{j})(n\overline{m}%
F_{j})+\phi _{0}(m\overline{m}\partial \ell )+\phi _{1}(m\overline{m}%
\partial m)]+ \\ 
\qquad +n^{\lambda }[(m\overline{m}F_{j})(\ell \overline{m}F_{j})+\overline{%
\phi _{\widetilde{0}}}(m\overline{m}\partial n)+\overline{\phi _{\widetilde{1%
}}}(m\overline{m}\partial m)]- \\ 
\qquad -m^{\lambda }[2(\ell \overline{m}F_{j})(n\overline{m}F_{j})+\phi
_{1}(\ell \overline{m}\partial m)+\overline{\phi _{\widetilde{1}}}(n%
\overline{m}\partial m)]\}=0 \\ 
\end{array}
\label{fe10}
\end{equation}

These last relations combined with the tetrad integrability conditions imply
relations between the surface geometric quantities and the gauge field
invariants. For that we will use the following relations of my spin
coefficients and the ordinary Newman-Penrose ones%
\begin{equation}
\begin{tabular}{|l|}
\hline
$\alpha =\frac{1}{4}[(\ell n\partial \overline{m})+(\ell \overline{m}%
\partial n)-(n\overline{m}\partial \ell )-2(m\overline{m}\partial \overline{m%
})]$ \\ \hline
$\beta =\frac{1}{4}[(\ell n\partial m)+(\ell m\partial n)-(nm\partial \ell
)-2(m\overline{m}\partial m)]$ \\ \hline
$\gamma =\frac{1}{4}[(nm\partial \overline{m})-(n\overline{m}\partial m)-(m%
\overline{m}\partial n)+2(\ell n\partial n)]$ \\ \hline
$\varepsilon =\frac{1}{4}[(\ell m\partial \overline{m})-(\ell \overline{m}%
\partial m)-(m\overline{m}\partial \ell )+2(\ell n\partial \ell )]$ \\ \hline
$\mu =-\frac{1}{2}[(m\overline{m}\partial n)+(nm\partial \overline{m})+(n%
\overline{m}\partial m)]$ \\ \hline
$\pi =\frac{1}{2}[(\ell n\partial \overline{m})-(n\overline{m}\partial \ell
)-(\ell \overline{m}\partial n)]$ \\ \hline
$\rho =\frac{1}{2}[(\ell \overline{m}\partial m)+(\ell m\partial \overline{m}%
)-(m\overline{m}\partial \ell )]$ \\ \hline
$\tau =\frac{1}{2}[(nm\partial \ell )+(\ell m\partial n)+(\ell n\partial m)]$
\\ \hline
$\kappa =(\ell m\partial \ell )\quad ,\quad \sigma =(\ell m\partial m)$ \\ 
\hline
$\nu =-(n\overline{m}\partial n)\quad ,\quad \lambda =-(n\overline{m}%
\partial \overline{m})$ \\ \hline
\end{tabular}
\label{fe11}
\end{equation}%
and the inverse relations 
\begin{equation}
\begin{array}{l}
(\ell n\partial \ell )=\varepsilon +\overline{\varepsilon }\quad ,\quad
(\ell m\partial \ell )=\kappa \quad ,\quad (nm\partial \ell )=\tau -%
\overline{\alpha }-\beta  \\ 
(\ell n\partial n)=\gamma +\overline{\gamma }\quad ,\quad (\ell m\partial n)=%
\overline{\alpha }+\beta -\overline{\pi }\quad ,\quad (nm\partial n)=-%
\overline{\nu } \\ 
(\ell n\partial m)=\tau +\overline{\pi }\quad ,\quad (\ell m\partial
m)=\sigma \quad ,\quad (\ell \overline{m}\partial m)=\overline{\varepsilon }%
-\varepsilon +\rho  \\ 
(nm\partial m)=-\overline{\lambda }\quad ,\quad (n\overline{m}\partial m)=%
\overline{\gamma }-\gamma -\overline{\mu }\quad ,\quad (m\overline{m}%
\partial m)=\overline{\alpha }-\beta  \\ 
(m\overline{m}\partial \ell )=\overline{\rho }-\rho \quad ,\quad (m\overline{%
m}\partial n)=\overline{\mu }-\mu  \\ 
\\ 
\nabla _{\mu }\ell ^{\mu }=\varepsilon +\overline{\varepsilon }-\rho -%
\overline{\rho }\quad ,\quad \nabla _{\mu }n^{\mu }=\mu +\overline{\mu }%
-\gamma -\overline{\gamma } \\ 
\nabla _{\mu }m^{\mu }=\overline{\pi }+\beta -\tau -\overline{\alpha } \\ 
\end{array}
\label{fe12}
\end{equation}%
which are implied by the following formula\cite{CHAND} of the covariant
derivatives of the null tetrad 
\begin{equation}
\begin{array}{l}
\nabla _{\mu }\ell _{\nu }=(\gamma +\overline{\gamma })\ell _{\mu }\ell
_{\nu }-\overline{\tau }\ell _{\mu }m_{\nu }-\tau \ell _{\mu }\overline{m}%
_{\nu }+(\varepsilon +\overline{\varepsilon })n_{\mu }\ell _{\nu }- \\ 
\qquad -\overline{\kappa }n_{\mu }m_{\nu }-\kappa n_{\mu }\overline{m}_{\nu
}-(\alpha +\overline{\beta })m_{\mu }\ell _{\nu }+\overline{\sigma }m_{\mu
}m_{\nu }+ \\ 
\qquad +\rho m_{\mu }\overline{m}_{\nu }-(\overline{\alpha }+\beta )%
\overline{m}_{\mu }\ell _{\nu }+\overline{\rho }\overline{m}_{\mu }m_{\nu
}+\sigma \overline{m}_{\mu }\overline{m}_{\nu } \\ 
\\ 
\nabla _{\mu }n_{\nu }=-(\gamma +\overline{\gamma })\ell _{\mu }n_{\nu }+\nu
\ell _{\mu }m_{\nu }+\overline{\nu }\ell _{\mu }\overline{m}_{\nu
}-(\varepsilon +\overline{\varepsilon })n_{\mu }n_{\nu }+ \\ 
\qquad +\pi n_{\mu }m_{\nu }+\overline{\pi }n_{\mu }\overline{m}_{\nu
}+(\alpha +\overline{\beta })m_{\mu }n_{\nu }-\lambda m_{\mu }m_{\nu }- \\ 
\qquad -\overline{\mu }m_{\mu }\overline{m}_{\nu }+(\overline{\alpha }+\beta
)\overline{m}_{\mu }n_{\nu }-\mu \overline{m}_{\mu }m_{\nu }-\overline{%
\lambda }\overline{m}_{\mu }\overline{m}_{\nu } \\ 
\\ 
\nabla _{\mu }m_{\nu }=\overline{\nu }\ell _{\mu }\ell _{\nu }-\tau \ell
_{\mu }n_{\nu }+(\gamma -\overline{\gamma })\ell _{\mu }m_{\nu }+\overline{%
\pi }n_{\mu }\ell _{\nu }-\kappa n_{\mu }n_{\nu }+ \\ 
\qquad +(\varepsilon -\overline{\varepsilon })n_{\mu }m_{\nu }-\overline{\mu 
}m_{\mu }\ell _{\nu }+\rho m_{\mu }n_{\nu }+(\overline{\beta }-\alpha
)m_{\mu }m_{\nu }- \\ 
\qquad -\overline{\lambda }\overline{m}_{\mu }\ell _{\nu }+\sigma \overline{m%
}_{\mu }n_{\nu }+(\overline{\alpha }-\beta )\overline{m}_{\mu }m_{\nu } \\ 
\end{array}
\label{fe13}
\end{equation}

The field equations (\ref{fe5}) become 
\begin{equation}
\begin{array}{l}
m^{\mu }\partial _{\mu }\phi _{0}+\overline{m}^{\mu }\partial _{\mu }%
\overline{\phi _{0}}+\phi _{0}[3\beta -2\tau +\overline{\alpha }]+\overline{%
\phi _{0}}[3\overline{\beta }-2\overline{\tau }+\alpha ]+ \\ 
\qquad +2(nmF_{j})(n\overline{m}F_{j})=0 \\ 
\\ 
\ell ^{\mu }\partial _{\mu }\phi _{0}+\phi _{0}[3\varepsilon +\overline{%
\varepsilon }-\rho ]+\phi _{1}[\tau +\overline{\pi }]+(\ell nF_{j})(n%
\overline{m}F_{j})=0%
\end{array}
\label{fe14}
\end{equation}%
The field equations (\ref{fe7}) become 
\begin{equation}
\begin{array}{l}
\overline{m}^{\mu }\partial _{\mu }\phi _{\widetilde{0}}+m^{\mu }\partial
_{\mu }\overline{\phi _{\widetilde{0}}}+\phi _{\widetilde{0}}[-3\alpha +2\pi
-\overline{\beta }]+\overline{\phi _{\widetilde{0}}}[-3\overline{\alpha }+2%
\overline{\pi }-\beta ]- \\ 
\qquad -2(\ell mF_{j})(\ell \overline{m}F_{j})=0 \\ 
\\ 
n^{\mu }\partial _{\mu }\phi _{\widetilde{0}}+\phi _{\widetilde{0}}[-3\gamma
-\overline{\gamma }+\mu ]-\phi _{\widetilde{1}}[\overline{\tau }+\pi ]-(\ell
nF_{j})(\ell mF_{j})=0 \\ 
\end{array}
\label{fe15}
\end{equation}%
The field equations (\ref{fe9}) become 
\begin{equation}
\begin{array}{l}
m^{\mu }\partial _{\mu }\overline{\phi _{\widetilde{1}}}+\overline{\phi _{%
\widetilde{1}}}[-3\overline{\alpha }+\beta +\overline{\pi }]+\overline{\phi
_{\widetilde{0}}}[\mu -\overline{\mu }]-(\ell \overline{m}F_{j})(m\overline{m%
}F_{j})=0 \\ 
\\ 
m^{\mu }\partial _{\mu }\phi _{1}+\phi _{1}[3\beta -\overline{\alpha }-\tau
]+\phi _{0}[\rho -\overline{\rho }]-(n\overline{m}F_{j})(m\overline{m}%
F_{j})=0 \\ 
\\ 
\ell ^{\mu }\partial _{\mu }\phi _{1}+n^{\mu }\partial _{\mu }\overline{\phi
_{\widetilde{1}}}+\phi _{1}[3\varepsilon -2\rho -\overline{\varepsilon }]+%
\overline{\phi _{\widetilde{1}}}[-3\overline{\gamma }+2\overline{\mu }%
+\gamma ]- \\ 
\qquad -2(\ell \overline{m}F_{j})(n\overline{m}F_{j})=0 \\ 
\end{array}
\label{fe16}
\end{equation}

Using the Newman-Penrose spin coefficients, the field equations (\ref{fe2})
become 
\begin{equation}
\begin{array}{l}
m^{\mu }D_{\mu }(\ell \overline{m}F_{j})+\overline{m}^{\mu }D_{\mu }(\ell
mF_{j})+(\ell \overline{m}F_{j})[\overline{\pi }-2\overline{\alpha }]+ \\ 
\qquad +(\ell mF_{j})[\pi -2\alpha ]=0 \\ 
\\ 
m^{\mu }D_{\mu }(n\overline{m}F_{j})+\overline{m}^{\mu }D_{\mu }(nmF_{j})+(n%
\overline{m}F_{j})[2\beta -\tau ]+ \\ 
\qquad +(nmF_{j})[2\overline{\beta }-\overline{\tau }]=0 \\ 
\\ 
\ell ^{\mu }D_{\mu }(nmF_{j})+n^{\mu }D_{\mu }(\ell mF_{j})+(nmF_{j})[2%
\overline{\varepsilon }-\overline{\rho }]]+ \\ 
\qquad +(\ell mF_{j})[\mu -2\gamma ]=0 \\ 
\end{array}
\label{fe17}
\end{equation}%
Their integrability conditions are satisfied identically.

The integrability condition of the equations (\ref{fe15}) is 
\begin{equation}
\begin{array}{l}
m^{\mu }\partial _{\mu }[(\ell nF_{j})(n\overline{m}F_{j})]+\overline{m}%
^{\mu }\partial _{\mu }[(\ell nF_{j})(nmF_{j})]-2\ell ^{\mu }\partial _{\mu
}[(nmF_{j})(n\overline{m}F_{j})]+ \\ 
\qquad +(2\beta +\overline{\pi }-2\tau )(\ell nF_{j})(n\overline{m}F_{j})+(2%
\overline{\beta }+\pi -2\overline{\tau })(\ell nF_{j})(nmF_{j})+ \\ 
\qquad +(\tau +\overline{\pi })(m\overline{m}F_{j})(n\overline{m}F_{j})-(%
\overline{\tau }+\pi )(m\overline{m}F_{j})(nmF_{j})+ \\ 
\qquad +2(\rho +\overline{\rho }-2\varepsilon -2\overline{\varepsilon }%
)(nmF_{j})(n\overline{m}F_{j})=0 \\ 
\end{array}
\label{fe18}
\end{equation}%
where the tetrad commutation relations\cite{CHAND} are used. The equations (%
\ref{fe16}) imply 
\begin{equation}
\begin{array}{l}
m^{\mu }\partial _{\mu }[(\ell nF_{j})(\ell \overline{m}F_{j})]+\overline{m}%
^{\mu }\partial _{\mu }[(\ell nF_{j})(\ell mF_{j})]-2n^{\mu }\partial _{\mu
}[(\ell mF_{j})(\ell \overline{m}F_{j})]+ \\ 
\qquad +(-2\overline{\alpha }+2\overline{\pi }-\tau )(\ell nF_{j})(\ell 
\overline{m}F_{j})+(-2\alpha +2\pi -\overline{\tau })(\ell nF_{j})(\ell
mF_{j})+ \\ 
\qquad +(\tau +\overline{\pi })(m\overline{m}F_{j})(\ell \overline{m}F_{j})-(%
\overline{\tau }+\pi )(m\overline{m}F_{j})(\ell mF_{j})+ \\ 
\qquad +2(2\gamma +2\overline{\gamma }-\mu -\overline{\mu })(\ell
mF_{j})(\ell \overline{m}F_{j})=0 \\ 
\end{array}
\label{fe19}
\end{equation}%
and the equations (\ref{fe17}) imply the integrability condition 
\begin{equation}
\begin{array}{l}
\ell ^{\mu }\partial _{\mu }[(n\overline{m}F_{j})(m\overline{m}%
F_{j})]+n^{\mu }\partial _{\mu }[(\ell \overline{m}F_{j})(m\overline{m}%
F_{j})]-2m^{\mu }\partial _{\mu }[(\ell \overline{m}F_{j})(n\overline{m}%
F_{j})]+ \\ 
\qquad +(2\varepsilon -2\rho -\overline{\rho })(n\overline{m}F_{j})(m%
\overline{m}F_{j})+(2\overline{\mu }-2\overline{\gamma }+\mu )(\ell 
\overline{m}F_{j})(m\overline{m}F_{j})+ \\ 
\qquad +(\rho -\overline{\rho })(\ell nF_{j})(n\overline{m}F_{j})+(\overline{%
\mu }-\mu )(\ell nF_{j})(\ell \overline{m}F_{j})+ \\ 
\qquad +2(2\overline{\alpha }-2\beta +\tau -\overline{\pi })(\ell \overline{m%
}F_{j})(n\overline{m}F_{j})=0 \\ 
\end{array}
\label{fe20}
\end{equation}%
Notice that the curvature terms cancel out in all these integrability
conditions.

The above integrability conditions are the null tetrad forms of the
following relations implied by the gauge field equations (\ref{fe1}). 
\begin{equation}
\begin{array}{l}
\nabla _{\mu }\{\Gamma ^{\mu \lambda \rho \sigma }F_{j\nu \lambda }F_{j\rho
\sigma }-\frac{1}{4}\delta _{\ \nu }^{\mu }(\Gamma ^{\tau \lambda \rho
\sigma }F_{j\tau \lambda }F_{j\rho \sigma })\}=-\frac{1}{4}(\nabla _{\nu
}\Gamma ^{\tau \lambda \rho \sigma })F_{j\tau \lambda }F_{j\rho \sigma } \\ 
\\ 
\Gamma ^{\mu \nu \rho \sigma }=\frac{1}{2}[(\ell ^{\mu }m^{\nu }-\ell ^{\nu
}m^{\mu })(n^{\rho }\overline{m}^{\sigma }-n^{\sigma }\overline{m}^{\rho
})+(n^{\mu }\overline{m}^{\nu }-n^{\nu }\overline{m}^{\mu })(\ell ^{\rho
}m^{\sigma }-\ell ^{\sigma }m^{\rho })+c.c.] \\ 
\end{array}
\label{fe21}
\end{equation}%
which takes the following form with ordinary derivatives derivatives 
\begin{equation}
\begin{array}{l}
\frac{1}{\sqrt{-g}}\partial _{\mu }\{\sqrt{-g}[\Gamma ^{\mu \lambda \rho
\sigma }F_{j\nu \lambda }F_{j\rho \sigma }-\frac{1}{4}\delta _{\ \nu }^{\mu
}(\Gamma ^{\tau \lambda \rho \sigma }F_{j\tau \lambda }F_{j\rho \sigma })]\}=
\\ 
\qquad =-\frac{1}{4\sqrt{-g}}\partial _{\nu }(\sqrt{-g}\Gamma ^{\tau \lambda
\rho \sigma })F_{j\tau \lambda }F_{j\rho \sigma } \\ 
\end{array}
\label{fe22}
\end{equation}%
It is well-known that the translation generators of a generally covariant
Lagrangian are first class constaints which must vanish. But notice that the
left term of the above equation looks like the energy-momentum tensor of the
new action. It is not generally conserved, but if the null tetrad is
time-independent, it does provide a conserved energy for the gauge field in
an external spacetime complex structure. In fact this energy was used (\ref%
{i15}) and we will use below, in the next section.

\section{THE ENERGY OF TWO EXTERNAL "COLORED" SOURCES}

\setcounter{equation}{0}

I will first consider the energy of two electric charges located at $%
\overrightarrow{x}_{1}$\ and $\overrightarrow{x}_{2}$. The\ particle and
electromagnetic energy momentum tensors are 
\begin{equation}
\begin{array}{l}
T_{(p)\ \nu }^{\mu }=\underset{n}{\tsum }\frac{dx_{n}^{\mu }}{dt}P_{n\nu
}\delta ^{3}(\overrightarrow{x}-\overrightarrow{x}_{n}) \\ 
T_{(e)\ \nu }^{\mu }=-F^{\mu \rho }F_{\nu \rho }+\frac{1}{4}\delta _{\nu
}^{\mu }F^{\rho \sigma }F_{\rho \sigma } \\ 
\end{array}
\label{e1}
\end{equation}%
We know that the total energy momentum tensor is conserved, because of the
following EM field equations and the Lorentz force 
\begin{equation}
\begin{array}{l}
\partial _{\mu }F^{\mu \rho }=J^{\rho } \\ 
\frac{dP_{n}^{\mu }}{dt}=q_{n}F_{\ \nu }^{\mu }\frac{dx_{n}^{\nu }}{dt}%
\end{array}
\label{e2}
\end{equation}%
where the EM charge current is 
\begin{equation}
\begin{array}{l}
J^{\mu }=\underset{n}{\tsum }q_{n}\delta ^{3}(\overrightarrow{x}-%
\overrightarrow{x}_{n})\frac{dx_{n}^{\mu }}{dt} \\ 
\end{array}
\label{e3}
\end{equation}

The total energy of the two charges is 
\begin{equation}
\begin{array}{l}
E=\int d^{3}x\{[m_{1}\delta ^{3}(\overrightarrow{x}-\overrightarrow{x}%
_{1})+m_{2}\delta ^{3}(\overrightarrow{x}-\overrightarrow{x}_{2})]+ \\ 
\quad +[(-\partial _{i}A_{0})^{2}+\frac{1}{4}(-2(\partial _{i}A_{0})^{2})]\}=
\\ 
\quad =m_{1}+m_{2}+\frac{1}{2}\int d^{3}x(\partial _{i}A_{0})^{2} \\ 
\end{array}
\label{e4}
\end{equation}%
From the FE and the static current we find 
\begin{equation}
\begin{array}{l}
A_{0}=\frac{1}{4\pi }\frac{q_{1}}{\mid \overrightarrow{x}-\overrightarrow{x}%
_{1}\mid }+\frac{1}{4\pi }\frac{q_{2}}{\mid \overrightarrow{x}-%
\overrightarrow{x}_{2}\mid } \\ 
\end{array}
\label{e5}
\end{equation}%
Which gives 
\begin{equation}
\begin{array}{l}
E=m_{1}+m_{2}+\frac{q_{1}q_{2}}{\mid \overrightarrow{x}_{1}-\overrightarrow{x%
}_{2}\mid }+self-interaction \\ 
\end{array}
\label{e6}
\end{equation}%
Notice that the energy of two opposite charges has an upper bound $%
m_{1}+m_{2}$ when their distance goes to infinity.

We will now compute the energy of two "gluonic" charges located at $%
\overrightarrow{x}_{1}=(0,0,d)$\ and $\overrightarrow{x}_{2}=(0,0,-d)$.
Because of the cylindrical symmetry of the system, we assume that its
complex structure may be asymptotically approximated by the cylindrical null
tetrad 
\begin{equation}
\begin{array}{l}
\ell _{\mu }dx^{\mu }=dt-dz \\ 
n_{\mu }dx^{\mu }=\frac{1}{2}\ (dt+dz) \\ 
m_{\mu }dx^{\mu }=\frac{-1}{\sqrt{2}}\ (d\rho +i\rho d\varphi ) \\ 
\sqrt{-g}=\rho%
\end{array}
\label{e7}
\end{equation}%
with its contravariant coordinates 
\begin{equation}
\begin{array}{l}
\ell ^{\mu }\partial _{\mu }=\partial _{t}+\partial _{z} \\ 
n^{\mu }\partial _{\mu }=\frac{1}{2}\ (\partial _{t}-\partial _{z}) \\ 
m^{\mu }\partial _{\mu }=\frac{1}{\sqrt{2}}\ (\partial _{\rho }+\frac{i}{%
\rho }\partial _{\varphi })%
\end{array}
\label{e8}
\end{equation}

The gluonic energy current is 
\begin{equation}
\begin{array}{l}
T_{(g)\ 0}^{\mu }=\Gamma ^{\mu \tau \rho \sigma }F_{0\tau }F_{\rho \sigma }-%
\frac{1}{4}\delta _{0}^{\mu }\Gamma ^{\nu \tau \rho \sigma }F_{\nu \tau
}F_{\rho \sigma } \\ 
\end{array}
\label{e9}
\end{equation}%
which is conserved for any static complex structure. The conservation of the
total energy current is implied by the following gluonic FE and particle
energy relation ("Lorentz" force) 
\begin{equation}
\begin{array}{l}
\frac{1}{\sqrt{-g}}\partial _{\mu }[\sqrt{-g}\Gamma ^{\mu \nu \rho \sigma
}F_{\rho \sigma }]=-J_{(g)}^{\nu } \\ 
\\ 
\partial _{\mu }T_{(p)\ 0}^{\mu }=\sqrt{-g}F_{0\nu }J_{(g)}^{\nu }%
\end{array}
\label{e10}
\end{equation}%
where I assume a $U(1)$ gauge group for convenience and the "gluonic"
current 
\begin{equation}
\begin{array}{l}
J_{g}^{0}=J_{g}^{1}=J_{g}^{2}=0 \\ 
J_{g}^{3}=\Phi (\rho )[q_{1}\delta (z-d)+q_{2}\delta (z+d)] \\ 
with\quad \Phi (\rho )=\func{real}\quad and\quad \tint d\rho \Phi (\rho )=1
\\ 
\partial _{\mu }[\sqrt{-g}J_{g}^{\mu }]=0%
\end{array}
\label{e11}
\end{equation}%
in ($t,z,\rho ,\varphi $) coordinates. Notice that this precise form of
external current satisfies the relation 
\begin{equation}
\begin{array}{l}
\partial _{\mu }T_{(p)\ 0}^{\mu }=\sqrt{-g}F_{0\nu }J_{(g)}^{\nu }=0%
\end{array}
\label{e11a}
\end{equation}
and it is found to be confined.

We look for $t,\varphi $-independent solutions with $A_{0}=A_{1}=A_{2}=0$.
Then 
\begin{equation}
\begin{array}{l}
(nmF)=-\frac{1}{2}\partial _{z}(m^{\mu }A_{\mu })\quad ,\quad (\ell
mF)=\partial _{z}(m^{\mu }A_{\mu }) \\ 
\frac{\partial ^{2}}{\partial z^{2}}(m^{\mu }A_{\mu })=-(m_{\nu }J_{g}^{\nu
}) \\ 
\partial _{z}\partial _{\rho }\{\rho \lbrack (m^{\mu }A_{\mu })+(\overline{m}%
^{\mu }A_{\mu })]\}=0%
\end{array}
\label{e12}
\end{equation}%
We find the solution 
\begin{equation}
\begin{array}{l}
(nmF)=-\frac{1}{2}\partial _{z}(m^{\mu }A_{\mu })\quad ,\quad (\ell
mF)=\partial _{z}(m^{\mu }A_{\mu }) \\ 
(m^{\mu }A_{\mu })=\frac{i\rho \Phi (\rho )}{\sqrt{2}}%
[q_{1}G(z-d)+q_{2}G(z+d)] \\ 
\end{array}
\label{e13}
\end{equation}%
where the Green function $G(z,z^{\prime })$ satisfies the equation $\partial
_{z}^{2}G(z,z^{\prime })=\delta (z-z^{\prime })$. Its form is known to be 
\begin{equation}
\begin{array}{l}
G(z,z^{\prime })=\langle _{\frac{z^{\prime }+L}{2L}(z-L)\quad ,\quad
z^{\prime }\leq z\leq L}^{\frac{z^{\prime }-L}{2L}(z+L)\quad ,\quad -L\leq
z\leq z^{\prime }} \\ 
\end{array}
\label{e14}
\end{equation}%
where $(-L\ ,\ L)$ is a box for convenience.

For two opposite charges the energy of the system is 
\begin{equation}
\begin{array}{l}
E=m_{1}+m_{2}+\int dzd\rho d\varphi \ \rho \lbrack 2(nmF)(n\overline{m}F)+%
\frac{1}{2}(\ell mF)(\ell \overline{m}F)]= \\ 
\quad =m_{1}+m_{2}+2\pi \int dzd\rho \ \rho \lbrack (m^{\mu }A_{\mu })(%
\overline{m}_{\nu }J_{g}^{\nu })]= \\ 
\quad =m_{1}+m_{2}+2\pi q^{2}d\ [\int_{0}^{\infty }d\rho \ \rho ^{3}\Phi
^{2}(\rho )]+self-int+O(L)%
\end{array}
\label{e15}
\end{equation}

We find that the interaction energy of the two opposite charges is linearly
increasing with their distance. Notice that the present "gluonic" current is
not analogous to the EM current. It has only the $\varphi $-component. It
rotates around the $z$-axis at the $z$-positions of the particles.

\section{DISCUSSION}

\setcounter{equation}{0}

Confinement of the hadronic constituents (partons) is an experimental fact.
It is generally believed that non-Abelian gauge theories are confined, but
it has not yet been mathematically proven. In four dimensions up to now,
linear perturbative potentials have been derived from Lagrangian models with
second order derivatives, which have serious unitarity problems. As far as I
know, the present Lagrangian with first order derivatives is the unique four
dimensional model which exhibits confinement of external sources. But the
characteristic property of the model, its independence on the metric of the
spacetime, which assures its renormalizability, imposes severe constraints
on the permitted Lagrangian terms.

Despite my efforts, I have not yet found a way to incorporate terms with the
Dirac field. Therefore the "external sources" cannot be classical
considerations of the well known fermionic currents. It seems that these
"external sources" have to emerge from the solitonic sectors of the model.

The model has a quite rich solitonic sector\cite{RAG2008a}. It is well known
that spacetimes with two geodetic and shear free null congruences, which
admit an integrable Lorentzian complex structure, exhibit fermionic
solitonic properties. A typical example is the Kerr-Newman spacetime which
has\cite{CART},\cite{NEWM} the fermionic gyromagnetic ratio g=2. These
fermionic solitons appear without any Dirac field present in the action.
Therefore it is natural to consider that the confined "external sources" of
the present work may emerge as gauge field excitations of these fermionic
solitons.

\bigskip

\end{document}